\documentclass[10pt,conference]{IEEEtran}
\IEEEoverridecommandlockouts
\IEEEsettopmargin{t}{0.9in}
\IEEEsettextheight{0.9in}{1.05in}
\usepackage[T1]{fontenc}
\usepackage{cite}
\usepackage{comment}
\usepackage{amsmath,amssymb,amsfonts}
\usepackage{algorithmic}
\usepackage{graphicx}
\usepackage{float}
\usepackage{textcomp}
\usepackage{xcolor}
\usepackage{tikz}
\usetikzlibrary{arrows.meta, positioning, fit, calc, backgrounds}
\usepackage{siunitx}
\usepackage{caption}
\usepackage[nolist]{acronym}
\usepackage{url}
\usepackage{subcaption}
\usepackage{booktabs}
\usepackage[bookmarks=false, hidelinks, draft]{hyperref}

\renewcommand{\thetable}{\arabic{table}}

\newcommand{\mycomment}[1]{}

\def\BibTeX{{\rm B\kern-.05em{\sc i\kern-.025em b}\kern-.08em
    T\kern-.1667em\lower.7ex\hbox{E}\kern-.125emX}}

\newcommand\copyrightnotice{%
\begin{tikzpicture}[remember picture,overlay]
\node[anchor=south,yshift=10pt] at (current page.south) {%
\footnotesize \parbox{0.9\textwidth}{\centering
Accepted to be published in: 2026 IEEE 37th International Symposium on Personal, Indoor and Mobile Radio Communications (PIMRC), Singapore, Sept. 1--4, 2026. \\
\textcopyright~2026 IEEE. Personal use of this material is permitted.
Permission from IEEE must be obtained for all other uses, in any current or
future media, including reprinting/republishing this material for advertising
or promotional purposes, creating new collective works, for resale or
redistribution to servers or lists, or reuse of any copyrighted component
of this work in other works.}};
\end{tikzpicture}}

\begin{document}
\bstctlcite{IEEEexample:BSTcontrol}

\title{Multi-Agent Reinforcement Learning for Base Station Placement in TDOA-Based Localization}
\author{\IEEEauthorblockN{Bastian Perner, Pratik Gajanan Raut, Maximilian Lübke, and Norman Franchi}\\
\IEEEauthorblockA{\textit{Institute for Smart Electronics and Systems}\\
\textit{Friedrich-Alexander-Universität Erlangen-Nürnberg}\\
Email: \{bastian.perner, pratik.pr.raut, maximilian.luebke, norman.franchi\}@fau.de}
}

\maketitle
\copyrightnotice

\begin{acronym}

\acro{snr}[SNR]{Signal-to-Noise Ratio}
\acro{mdp}[MDP]{Markov Decision Process}
\acro{bs}[BS]{base station}
\acro{cir}[CIR]{Channel Impulse Response}
\acro{drl}[DRL]{Deep Reinforcement Learning}
\acro{gdop}[GDOP]{Geometric Dilution of Precision}
\acro{gnss}[GNSS]{Global Navigation Satellite System}
\acro{los}[LOS]{Line of Sight}
\acro{mae}[MAE]{Mean Absolute Error}
\acro{mpc}[MPC]{Multipath Component}
\acro{marl}[MARL]{Multi-Agent Reinforcement Learning}
\acro{nlos}[NLOS]{Non-Line of Sight}
\acro{npn}[NPN]{Non-Public Network}
\acro{ppo}[PPO]{Proximal Policy Optimization}
\acro{rl}[RL]{Reinforcement Learning}
\acro{tanh}[tanh]{hyperbolic tangent}
\acro{tdoa}[TDOA]{Time Difference of Arrival}
\acro{toa}[TOA]{Time of Arrival}
\acro{uav}[UAV]{Unmanned Aerial Vehicle}
\acro{ue}[UE]{User Equipment}
\acro{ls}[LS]{Least-Squares}
\acro{cnn}[CNN]{Convolutional Neural Network}
\acro{gae}[GAE]{Generalized Advantage Estimation}
\acro{ctde}[CTDE]{Centralized Training with Decentralized Execution}
\acro{ga}[GA]{Genetic Algorithm}
\acro{sa}[SA]{Simulated Annealing}
\acro{cdf}[CDF]{Cumulative Distribution Function}
\acro{ci}[CI]{confidence interval}

\end{acronym}

\begin{abstract}
  Accurate localization of devices is a key capability for emerging 5G and 6G networks 
and depends on effective \ac{bs} placement. Conventional geometry-based approaches such as \ac{gdop}
ignore realistic propagation effects such as \ac{nlos} shadowing and multipath-induced \ac{toa} bias caused by buildings.
This paper proposes a ray-tracing-assisted \ac{marl} framework for environment-aware \ac{bs} placement in 
\ac{tdoa} localization systems. \ac{ppo} agents are trained on \acp{cir} generated from a detailed 3D model 
of a university campus. Each agent cooperatively places one \ac{bs} while optimizing a shared reward 
that combines localization accuracy and coverage. The approach is evaluated on five campus segments
with varying propagation characteristics. Results show that the learned policy achieves localization
accuracy comparable to conventional \ac{gdop}-based placement, lowering the average localization \ac{mae}
by about 3\,\% relative to the stronger (mean-optimized) geometric baseline. The behavior is
segment-dependent, with a clear improvement on individual segments (up to about 14\,\%) and comparable
or slightly higher error on the others. These findings indicate that incorporating site-specific propagation data into
the placement process can match and selectively improve upon purely geometric strategies, motivating
further work toward consistent gains.
\end{abstract}

\acresetall


\section{Introduction}
    \label{sec:introduction}
    Precise localization is a key feature for 5G and 6G verticals such as
Industry 4.0, asset tracking, and \ac{npn}~\cite{wen2022private}.
As sensing and communication converge in 6G architectures,
network-based positioning becomes a core
requirement~\cite{shatov2023cooperative, shatov2026ai}. \ac{tdoa} is an
established method for network-based localization that requires no \ac{ue}
synchronization and directly exploits existing \ac{bs}
infrastructure~\cite{dwivedi2021positioning}. Among the design parameters
that impact localization accuracy, \ac{bs} placement plays a
dominant role~\cite{ordonezlucena2019npn}. \ac{gdop}, derived from the Fisher
Information Matrix, is the standard metric for quantifying this geometric
dependency~\cite{comuniello2020tdoa}. However, \ac{gdop} captures purely
geometric configuration and neglects propagation effects such as \ac{nlos}
propagation, multipath-induced \ac{toa} bias, and path
loss~\cite{wang2020dop}. As a result, localization accuracy deviates
systematically from \ac{gdop}-based predictions, and exhaustive search for
optimal \ac{bs} placement remains infeasible due to combinatorial
complexity $\mathcal{O}\!\left(\binom{M}{N}\right)$.

\ac{gdop}-based placement optimization is well established in \ac{gnss} and
radar networks~\cite{peral2018survey}, where predominantly \ac{los}
propagation permits purely geometric objective functions. These
approaches do not transfer to cellular deployments, where \ac{nlos}
components fundamentally change the effective measurement geometry. Heuristic
methods such as \ac{ga} and \ac{sa} have been applied to localization
infrastructure planning~\cite{diezgonzalez2020memetic}
and measurement location planning for radio environment
maps~\cite{perner2025optimized}, but exhibit limited scalability and
retain geometric objective functions, leaving the core problem unresolved. \ac{drl} has been explored for network planning targeting
coverage and throughput~\cite{liu2018uav} as well as for \ac{uav} \ac{bs}
positioning~\cite{qiu2020drl}, yet neither formulation addresses localization
accuracy as an optimization objective. The work in~\cite{li2021rl}
investigates \ac{rl}-based \ac{bs} placement with a \ac{tdoa}-based
objective, but relies on a simplified noise model and omits site-specific
propagation effects. Ray-tracing-based \ac{cir} data has been used to capture
the structure of \acp{mpc} relevant for \ac{tdoa}
localization~\cite{desousa2018raytracing}, without extending the framework
to placement optimization. To the best of our knowledge, no prior work has
combined \ac{rl} with ray-tracing-based propagation modeling to optimize
\ac{bs} placement for \ac{tdoa} localization accuracy.

This raises the question whether \ac{marl} can leverage propagation effects
to improve \ac{bs} placement for \ac{tdoa}-based localization compared to
purely geometric approaches based on \ac{gdop}. Since field measurements
at all candidate positions are infeasible, ray-tracing provides the only
practical means to obtain complete propagation data for every
\ac{bs}--\ac{ue} combination.

This work proposes a \ac{marl} framework in which agents cooperatively place
\acp{bs} on a discrete 3D candidate grid. Policy training is conducted on
\ac{cir} data generated by Sionna RT~\cite{hoydis2023sionna}, allowing the
learned policy to implicitly capture \ac{nlos} regions and multipath
structure without requiring an explicit propagation model at inference time.
The framework is evaluated on a realistic outdoor campus scenario with building-induced
\ac{nlos} conditions.

While prior approaches tackle geometry or propagation in isolation,
no existing method combines both for \ac{tdoa}-based \ac{bs} placement. This work addresses that gap
with the following contributions:
\begin{itemize}
    \item A cooperative \ac{marl} formulation for \ac{tdoa} \ac{bs}
    placement, where $N$ agents jointly optimize a shared reward
    combining localization accuracy and coverage.
    \item Environment-aware placement through ray-tracing-based
    \ac{cir} training, enabling the policy to account for \ac{nlos}
    and multipath effects.
    \item Empirical evaluation across campus segments with varying
    propagation complexity, showing localization accuracy comparable to
    \ac{gdop} baselines on average and a segment-dependent \ac{mae} reduction
    of up to about 14\% on individual segments.
\end{itemize}

The remainder of this paper is structured as follows.
Section~\ref{sec:system_model} introduces the signal model, \ac{tdoa}
localization, and the problem formulation. Section~\ref{sec:methodology}
describes the proposed method, followed by the experimental setup in
Section~\ref{sec:experimental_setup}. Section~\ref{sec:results} presents and
discusses the results, and Section~\ref{sec:conclusion} concludes the paper.
\section{System Model}
    \label{sec:system_model}
    This section introduces the signal model, derives the \ac{tdoa} localization procedure, 
and formulates the \ac{bs} placement optimization problem.

\subsection{Signal Model}

The wireless channel between a \ac{bs} and a \ac{ue} is characterized by its \ac{cir}, 
which describes the superposition of all propagation paths between the two locations. 
For \ac{bs}~$i$ and a given \ac{ue}, the \ac{cir} is modeled as a sum of $L_i$ \acp{mpc}:
\begin{equation}
    h_i(t) = \sum_{l=1}^{L_i} a_{i,l} \cdot \delta(t - \tau_{i,l}),
\end{equation}
where $a_{i,l}$ and $\tau_{i,l}$ denote the complex amplitude and propagation delay of the $l$-th path of \ac{bs} $i$, 
respectively. The \ac{toa} is estimated as the delay of the strongest received component:
\begin{equation}
    \hat{\tau}_i = \tau_{i,l^*}, \quad l^* = \arg\max_l |a_{i,l}|.
\end{equation}
Under \ac{los} conditions this estimate coincides with the direct-path delay, 
but it incurs a positive bias whenever the direct path is obstructed. In this work, 
\acp{cir} are obtained from deterministic ray-tracing, which models the site-specific \acp{mpc} including reflections, 
diffractions, and \ac{nlos} shadowing. The resulting \ac{nlos}-induced bias is thus captured in the data, 
unlike purely geometry-based metrics such as \ac{gdop}, which inherently ignore propagation effects.

\subsection{\ac{tdoa}-based Localization}

Given $N$ \acp{bs} at known positions $\mathbf{b}_i \in \mathbb{R}^3$ and a \ac{ue} 
at unknown position $\mathbf{p} \in \mathbb{R}^3$, the \ac{tdoa} measured between \ac{bs} $i$ and a reference \ac{bs} $1$ is:
\begin{equation}
    \Delta\tau_{i,1} = \frac{1}{c}\left(\|\mathbf{p} - \mathbf{b}_i\|_2 - \|\mathbf{p} - \mathbf{b}_1\|_2\right) + n_{i,1},
\end{equation}
where $c$ is the speed of light and $n_{i,1}$ denotes the measurement noise, comprising receiver thermal noise, 
synchronization errors between \acp{bs}, and \ac{nlos}-induced bias from obstructed propagation paths. 
Each observation constrains the \ac{ue} to a hyperboloid, and jointly solving $N-1$ such constraints yields 
the position estimate. For 2D localization with fixed \ac{ue} height, at least $N \geq 3$ \acp{bs} are required.

\begin{figure}[t]
\centering
\resizebox{\columnwidth}{!}{%
\begin{tikzpicture}[
    block/.style={
        rectangle, draw=black!70, fill=white,
        minimum height=2.6em, minimum width=6.2em,
        align=center, font=\small,
        rounded corners=2pt, line width=0.5pt,
        inner sep=4pt
    },
    agent/.style={
        rectangle, draw=black!60, fill=none,
        minimum height=2.2em, minimum width=4.6em,
        align=center, font=\footnotesize,
        rounded corners=1.5pt, line width=0.4pt,
        inner sep=3pt
    },
    envbox/.style={
        rectangle, draw=black!50, fill=gray!6,
        rounded corners=3pt, line width=0.6pt
    },
    agentbox/.style={
        rectangle, draw=blue!40, fill=blue!4,
        rounded corners=2pt, line width=0.5pt,
        inner sep=8pt
    },
    reward/.style={
        rectangle, draw=black!60, fill=orange!10,
        minimum height=2.6em, minimum width=6.2em,
        align=center, font=\footnotesize,
        rounded corners=1.5pt, line width=0.4pt,
        inner sep=4pt
    },
    arr/.style={-Stealth, line width=0.5pt, black!70},
    dataarr/.style={-Stealth, line width=0.5pt, black!50, dashed},
    lbl/.style={font=\sffamily\scriptsize, text=black!60}
]

\coordinate (boxL) at (-0.2, 0);
\coordinate (boxR) at (9.8, 0);

\node[block] (campus) at (0.8, 0)   {3D Campus\\Model};
\node[block] (sionna) at (4.8, 0)   {Sionna RT\\Ray Tracing};
\node[block] (cirdb)  at (8.8, 0)   {CIR / TOA\\Database};

\draw[arr] (campus) -- (sionna);
\draw[arr] (sionna) -- (cirdb);

\node[reward] (rew) at (0.8, -2.8) {Reward $r_t$};
\node[block]  (tdoa) at (8.8, -2.8) {TDOA\\LS Estimation};

\node[block, fill=blue!6, minimum width=8.5em] (policy) at (0.8, -4.8) {PPO Policy $\pi_\theta$};

\node[agent] (a1)    at (5.6, -4.8) {Agent\,1\\$\mathrm{BS}_1$};
\node[agent] (adots) at (7.4, -4.8) {$\cdots$};
\node[agent] (aN)    at (9.2, -4.8) {Agent\,$N$\\$\mathrm{BS}_N$};

\pgfdeclarelayer{middle}
\pgfsetlayers{background,middle,main}

\coordinate (anchorL) at (-1.0, 0);
\coordinate (anchorR) at (10.6, 0);

\begin{scope}[on background layer]
    \node[envbox, fill=green!4, draw=black!40,
          fit=(campus)(cirdb)(anchorL |- campus.north)(anchorR |- campus.south),
          inner sep=14pt,
          label={[lbl, anchor=north west]north west:Data Generation}] (databox) {};
    \node[envbox,
          fit=(rew)(tdoa)(policy)(aN)(anchorL |- policy.south)(anchorR |- rew.north),
          inner sep=14pt,
          label={[lbl, anchor=north west]north west:MARL Environment}] (env) {};
\end{scope}
\begin{pgfonlayer}{middle}
    \node[agentbox, fit=(a1)(adots)(aN),
          inner sep=8pt,
          label={[lbl, anchor=south east, yshift=-2pt]south east:shared weights}] (abox) {};
\end{pgfonlayer}

\draw[dataarr] (cirdb.south) -- (tdoa.north)
    node[lbl, midway, right, xshift=2pt] {CIR lookup};

\draw[arr] (rew.south) -- (policy.north)
    node[lbl, midway, right, xshift=2pt] {$r_t,\; s_{t+1}$};

\draw[arr] (policy.east) -- (abox.west)
    node[lbl, midway, above, yshift=1pt] {$a_t$};

\draw[arr] (abox.north -| tdoa.south) -- (tdoa.south)
    node[lbl, midway, right, xshift=2pt] {$p_1^t \!\ldots\! p_N^t$};

\draw[arr] (tdoa.west) -- (rew.east)
    node[lbl, midway, above, yshift=1pt] {MAE, cov};

\end{tikzpicture}%
}
\caption{System overview. Sionna~RT generates a \ac{cir} database from the 3D model. $N$~\ac{marl} agents cooperatively place \ac{bs} based on a shared reward combining \ac{tdoa} localization accuracy and coverage.}
\label{fig:system_overview}
\end{figure}
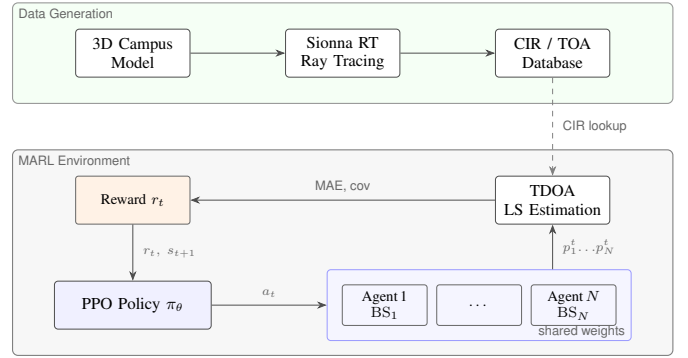

\subsection{Localization Error and Coverage}

The 2D position estimate $\hat{\mathbf{p}} = (\hat{x}, \hat{y})$ is obtained via \ac{ls} multilateration, 
and the localization error is the Euclidean distance to the true position:
\begin{equation}
    e = \|\hat{\mathbf{p}} - \mathbf{p}\|_2.
\end{equation}
Let $\mathcal{P}$ denote the set of all \ac{ue} positions in a segment. Not every \ac{ue} can be localized,
since detectability requires valid \ac{toa} measurements from at least three \acp{bs}. We denote the
detectable subset by $\mathcal{P}_d \subseteq \mathcal{P}$.
Non-detectable \acp{ue} are excluded from the \ac{mae} computation, and their fraction is reported as coverage separately,
capturing a distinct dimension of placement quality that \ac{mae} alone does not reflect.

\subsection{Problem Formulation}

Let $\mathcal{G} = \{g_1, \ldots, g_M\}$ be the set of $M$ discrete candidate \ac{bs} positions
on a 3D grid spanning the deployment area (the concrete grid dimensions and height levels are
specified in Section~\ref{sec:experimental_setup}). The task is to select
a subset $\mathcal{B} \subset \mathcal{G}$ of $N$ positions that jointly minimize localization error and non-coverage:
\begin{equation}
    \mathcal{B}^* = \arg\min_{\mathcal{B} \subset \mathcal{G}} \; f\!\left(\text{MAE}(\mathcal{B}),\, m(\mathcal{B});\, \beta\right),
\end{equation}
where $f:\mathbb{R}_{\ge 0}\times[0,1]\to\mathbb{R}$ is a scalarization of the two objectives that is
monotonically increasing in both arguments, so that higher localization error or larger non-coverage
yields a worse objective value, with $\beta \in [0,1]$ controlling their relative
weighting. Here,
\begin{equation}
    \text{MAE}(\mathcal{B}) = \frac{1}{|\mathcal{P}_d|} \sum_{\mathbf{p} \in \mathcal{P}_d} e(\mathbf{p}, \mathcal{B})
\end{equation}
is the mean localization error over the detectable
\acp{ue} $\mathcal{P}_d$, and $m(\mathcal{B}) = (|\mathcal{P}| - |\mathcal{P}_d|) / |\mathcal{P}|$ is
the non-coverage fraction. A concrete instantiation of $f$, used as the training reward, is
given in Section~\ref{sec:methodology}.
With $\mathcal{O}\!\left(\binom{M}{N}\right)$ possible subsets, exhaustive search becomes computationally 
infeasible even for moderate $M$ and $N$, motivating a learned optimization approach.
\section{Proposed Method}
    \label{sec:methodology}
    We formulate the \ac{bs} placement task as a sequential decision problem and solve it with cooperative \ac{marl}. 
This multi-agent formulation decomposes the joint placement of $N$ \acp{bs} into $N$ parallel decisions with shared learning. Each of the $N$ agents controls one \ac{bs}, 
and all agents share a common reward combining localization accuracy and coverage. 
The placement policy is optimized using \ac{ppo} with parameter sharing across agents. 
Fig.~\ref{fig:system_overview} presents the overall system architecture.

\subsection{MDP Formulation}
\label{subsec:mdp}
We model the \ac{bs} placement problem as a cooperative multi-agent
\ac{mdp} $\mathcal{M} = (\mathcal{S}, \mathcal{A}, \mathcal{T}, R, \gamma)$,
comprising the state space $\mathcal{S}$, the joint action space $\mathcal{A}$,
a deterministic transition function $\mathcal{T}$, a shared team reward $R$, and
the discount factor $\gamma$ (see Table~\ref{tab:hyperparams}). The $N$ agents act
simultaneously to optimize the joint placement of $N$ \acp{bs}. Each component is
detailed below.

\subsubsection{State space}
Let $\mathcal{G} \subset \mathbb{R}^3$ denote the set of placeable candidate
positions. The state at time step $t$ is the joint position of
all $N$ agents:
\begin{equation}
    s_t = \bigl(p_1^t, \ldots, p_N^t\bigr), \quad p_i^t \in \mathcal{G}.
\end{equation}
Agents are initialized at randomly sampled positions $p_i^0 \in \mathcal{G}$
without replacement.

\subsubsection{Action space}
Each agent $i$ maintains a current position $p_i^t \in \mathcal{G}$ at
time step $t$ and selects an action from its valid action set
\begin{equation}
    \mathcal{A}_i^t = \bigl\{ a \in \mathcal{N}(p_i^t) \mid a \notin
    \{p_j^t\}_{j \neq i} \bigr\},
\end{equation}
where $\mathcal{N}(p_i^t) \subseteq \mathcal{G}$ is a bounded set of local
relocations around $p_i^t$, comprising horizontal grid steps, height-level
changes, and the nearest unoccupied candidate positions. Targets occupied by
another agent are removed via action masking.

\subsubsection{Transition}
The transition function $\mathcal{T}$ is deterministic. All agents act
simultaneously based on a shared occupancy snapshot taken at the beginning
of each step. Each agent $i$ moves to its selected target position
$p_i' \in \mathcal{A}_i^t$. In case of conflicts, the agent with the
lowest index has priority while the remaining agents stay at their current position.
The subsequent state $s_{t+1} = (p_1^{t+1}, \ldots, p_N^{t+1})$ reflects
the updated agent locations.

\subsubsection{Reward}
All agents receive a shared team reward at each time step $t$:
\begin{equation}
    r_t = 2\left(\beta \cdot r_{\text{loc}} + (1-\beta) \cdot r_{\text{cov}}\right), \quad r_t \in [-2, 2],
\end{equation}
where $\beta \in [0,1]$ controls the trade-off between localization
accuracy and coverage, instantiating the general objective $f$ in Eq.~(5). Both components use a $\tanh$ mapping:
\begin{equation}
    r_{\text{loc}} = 1 - 2\tanh\!\left(\frac{\text{MAE}(\mathcal{B})}{\sigma_{\text{loc}}}\right), \quad
    r_{\text{cov}} = 1 - 2\tanh\!\left(\frac{m(\mathcal{B})}{\sigma_{\text{cov}}}\right),
\end{equation}
where $\text{MAE}(\mathcal{B})$ is the mean localization error and
$m(\mathcal{B})$ the fraction of \acp{ue} with insufficient \ac{toa}
measurements for the current \ac{bs} configuration $\mathcal{B}$.
The scaling parameters $\sigma_{\text{loc}}$ and $\sigma_{\text{cov}}$
control the sensitivity of each component.
The $\tanh$ mapping bounds each component to $[-1,1]$ and, through its saturating
shape, places the two heterogeneous quantities, namely the localization error and the coverage,
on a comparable scale while limiting the influence of
outliers on the gradient. Each component is a monotonically decreasing transform of $\text{MAE}(\mathcal{B})$ and $m(\mathcal{B})$, respectively,
so that maximizing $r_t$ realizes the general objective $f$ in Eq.~(5) under the weighting $\beta$.
Evaluating $r_t$ is dominated by one \ac{ls}-\ac{tdoa} solution per detectable
\ac{ue}, i.e., $\mathcal{O}(|\mathcal{P}|)$ per step.

\subsection{Network Architecture}

The policy network processes the voxelized grid state $s_t$ using a 3D \ac{cnn} backbone, 
which extracts spatial features from the occupancy representation of the environment.
For each agent, a region-of-interest of size $k{=}5$ voxels around its current position is extracted 
and mapped to a compact 64-dimensional token encoding both the local spatial context and the agent's position within the grid.

Each agent's token is passed to a shared actor head, which produces logits over the agent's local action set $\mathcal{A}_i^t$. 
Actions corresponding to occupied positions are suppressed via action masking before the softmax, 
ensuring that only valid placements are sampled. A centralized critic aggregates globally pooled features from all agents 
to produce a scalar value estimate, following the \ac{ctde} paradigm. All agents share the same network weights, 
reducing the number of trainable parameters and improving sample efficiency.

The policy is optimized using \ac{ppo} with a clipped surrogate objective and \ac{gae}.
Since all agents receive the same shared team reward,
the joint log-probability of all agents' actions is used during backpropagation.
With parameter sharing across agents and the centralized critic above, this
realizes the multi-agent PPO (MAPPO) scheme under the \ac{ctde} paradigm.
\ac{ppo} is preferred over deterministic-policy alternatives such as multi-agent deep deterministic policy gradient (MADDPG)
because its on-policy stochastic formulation natively supports the discrete,
mask-constrained action space and provides stable, sample-efficient updates in
the cooperative shared-reward setting.
\section{Experimental Setup}
    \label{sec:experimental_setup}
    We first describe the simulation environment and its segmentation,
then detail the \ac{rl} training configuration and the baseline methods used for comparison.

\begin{figure}[t]
    \centering
    \includegraphics[width=0.48\textwidth]{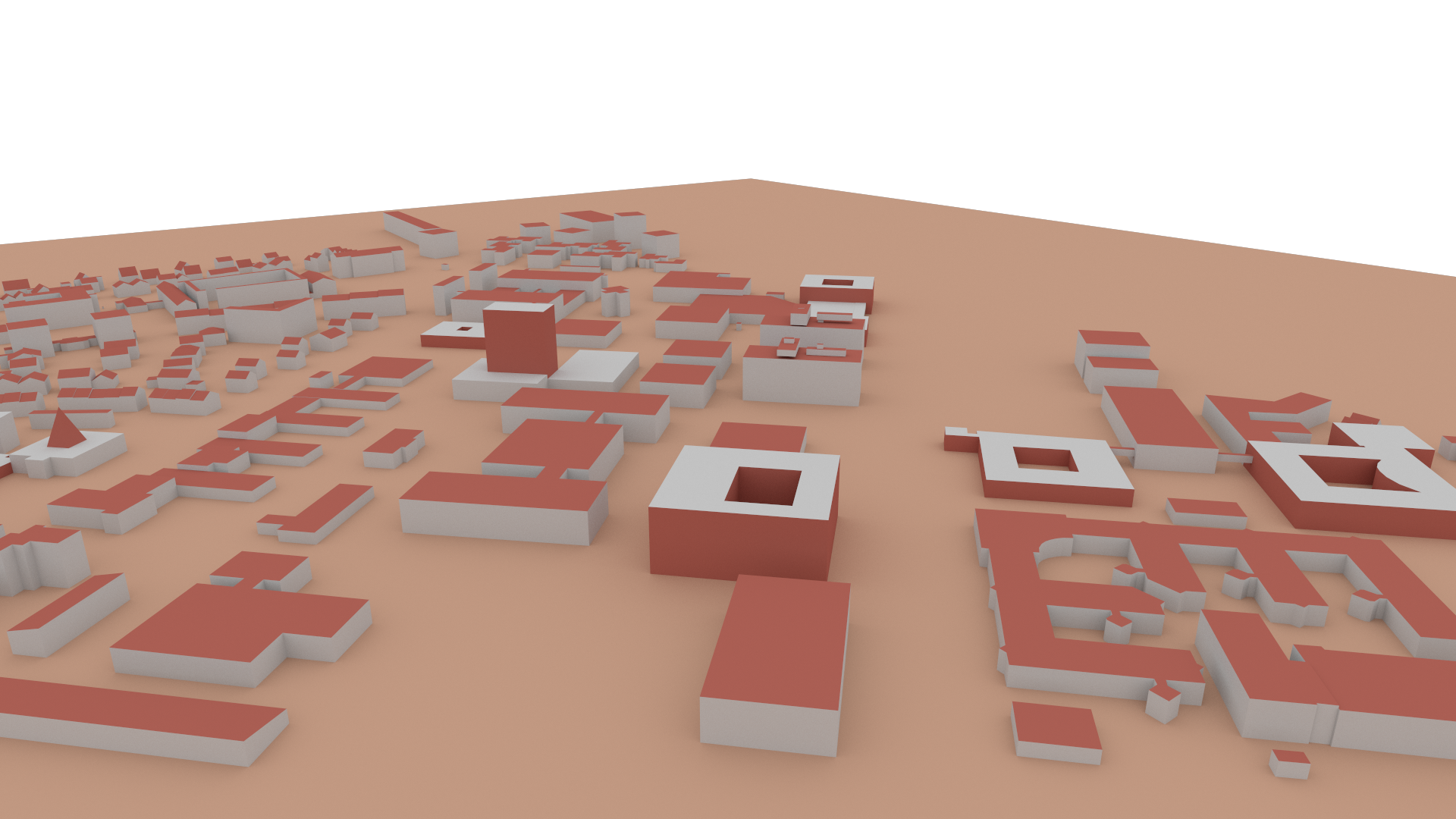}
    \caption{3D model of FAU Erlangen-Nürnberg Campus in Erlangen, Germany, created from OpenStreetMap data and imported into Sionna.}
    \label{fig:campus}
\end{figure}

\subsection{Scenario and Ray-Tracing Simulation}
Experiments are conducted on a model of the campus ``Technische Fakultät''
at FAU Erlangen-Nürnberg in Erlangen, Germany, shown in Fig.~\ref{fig:campus},
covering approximately $790\,\text{m} \times 1090\,\text{m}$.
Building geometries were extracted from OpenStreetMap\footnote{\copyright\ OpenStreetMap contributors.
Licensed under the Open Database License (ODbL) v1.0.} and imported into Sionna~\cite{hoydis2023sionna}
via a Blender model. Multiple multi-story buildings produce \ac{nlos} shadowing,
representative of a campus \ac{npn} deployment~\cite{wen2022private}. \acp{cir} are generated at
a carrier frequency of \SI{3.75}{\giga\hertz}, corresponding to the German campus network band,
for all combinations of $M \approx 300$ \ac{bs} candidate and $N_{\mathrm{UE}} \approx 350$ \ac{ue} positions per segment.
\ac{bs} candidates are arranged on an equidistant grid with \SI{20}{\meter} spacing
at three height levels (\SI{1}{\meter}, \SI{9}{\meter}, \SI{17}{\meter}). \acp{ue} occupy
a \SI{10}{\meter}-spaced grid at a fixed height of \SI{1.5}{\meter}. In both cases,
positions inside buildings are excluded. A sampling frequency of \SI{1}{\giga\hertz} provides
a delay resolution of \SI{1}{\nano\second}, supporting robust \ac{toa} extraction from
the \SI{100}{\mega\hertz} bandwidth-limited \ac{cir}. \ac{toa} estimates are obtained from
the strongest \ac{mpc}, which introduces a deterministic positive bias
under \ac{nlos} conditions~\cite{desousa2018raytracing}. \ac{ue} positions are
estimated via \ac{ls}-based \ac{tdoa} multilateration~\cite{dwivedi2021positioning},
requiring at least three \ac{toa} measurements per \ac{ue}.

\subsection{Environment Segmentation}

\begin{table}[t]
    \centering
    \caption{Selected campus segments and their characteristics. Vis.: average fraction of \ac{bs} candidates reachable from each \ac{ue} position.}
    \label{tab:segments}
    \begin{tabular}{cccc}
        \toprule
        Segment & \#\,UEs & \#\,BS Cand. & Vis.\ (\%) \\
        \midrule
        S4  & 300 & 244 & 67.8 \\
        S6  & 359 & 332 & 86.5 \\
        S9  & 313 & 290 & 74.3 \\
        S11 & 386 & 343 & 88.1 \\
        S14 & 423 & 338 & 93.3 \\
        \bottomrule
    \end{tabular}
\end{table}

The campus area is divided into $4 \times 4 = 16$ non-overlapping segments
of approximately $198\,\text{m} \times 273\,\text{m}$ each. 
This size ensures that each segment contains enough \ac{bs} candidates and \ac{ue} positions for
meaningful placement optimization, while also keeping
the combinatorial complexity of exhaustive baseline evaluation $\mathcal{O}\!\left(\binom{M}{N}\right)$
computationally feasible.
Each segment maintains its own \ac{bs} candidate grid,
\ac{ue} grid, and precomputed \ac{cir} lookup table. Five randomly selected segments (S3, S7, S8, S12, S16) 
serve as the training set and five as the evaluation set,
summarized in Table~\ref{tab:segments}. The evaluation segments cover a wide range of
propagation conditions, from heavily obstructed (S4, visibility 67.8\%) to predominantly open
environments (S14, visibility 93.3\%). Lower visibility implies more obstructed links between
\ac{bs} and \ac{ue} and thus stronger multipath-induced \ac{toa} bias.
As only building structures are modeled as obstacles, vegetation and street furniture are left for
future refinement. The training and evaluation sets do not overlap, allowing
to test generalization to unseen propagation conditions.

\subsection{Reinforcement Learning Configuration}
Throughout this work, $N = 5$ \acp{bs} are placed per segment. To improve generalization,
each training episode draws a segment randomly from the training set. The per-agent action space
is the bounded set of local relocations $\mathcal{N}(p_i^t)$ defined in Section~\ref{sec:methodology}, while invalid and occupied locations are excluded
via action masking. A checkpoint is saved every $50$ updates, and the model achieving
the highest $100$-episode moving average return is additionally saved as the best model per seed.
The reward weight $\beta = 0.5$ balances localization accuracy and coverage equally.
Each reward component is scaled to the range $[-1, 1]$ using a $\tanh$ mapping 
with scaling parameters $\sigma_\text{loc}$ and $\sigma_\text{cov}$.
Specifically, $\sigma_\text{loc} = 91\,\si{\meter}$ places the zero-crossing at approximately \SI{50}{\meter},
and $\sigma_\text{cov} = 0.5$ places it at a non-coverage fraction of $m \approx 0.27$,
corresponding to a coverage of approximately 73\%.
The \ac{ppo} hyperparameter configuration is given in Table~\ref{tab:hyperparams}.

\begin{table}[t]
\centering
\caption{PPO hyperparameter configuration.}
\label{tab:hyperparams}
\begin{tabular}{lc}
\toprule
Hyperparameter & Value \\
\midrule
Learning rate $\alpha$            & $10^{-4}$   \\
Discount factor $\gamma$          & $0.99$      \\
GAE parameter $\lambda$           & $0.95$      \\
Clip ratio $\epsilon$             & $0.20$      \\
Entropy coefficient $c_\text{ent}$ & $0.02$     \\
Value function coefficient $c_\text{vf}$ & $1.0$ \\
PPO epochs per update $K$         & $4$         \\
Mini-batch size                   & $256$       \\
Rollout length $T$                & $64$        \\
Max.\ gradient norm               & $0.5$       \\
Total training updates            & $10{,}000$ \\
\midrule
Reward weight $\beta$              & $0.5$   \\
Loc.\ scaling $\sigma_\text{loc}$  & $91$    \\
Cov.\ scaling $\sigma_\text{cov}$  & $0.5$   \\
\bottomrule
\end{tabular}
\end{table}

\subsection{Baselines}
As geometric baselines, we implement \ac{gdop} mean and median ranking, which select the
$N$ \ac{bs} candidate locations with the lowest average and median \ac{gdop} across
all \ac{ue} locations, respectively, for each segment. These baselines represent the geometric standard
for \ac{tdoa} localization and have been shown to outperform random and
coverage-based heuristics~\cite{comuniello2020tdoa,zhang2021dop,wang2020dop}.
All baselines use the same \ac{ue} and \ac{bs} candidate grids as the \ac{marl} agent.
Note that metaheuristic approaches such as \ac{ga} or \ac{sa} optimize the same geometric
\ac{gdop} objective and would thus converge to similar placements as the \ac{gdop} baselines
used here.

\subsection{Evaluation Metrics and Statistical Testing}
The primary evaluation metric is the \ac{mae} of the 2D position estimates across all \ac{ue} locations within a segment.
Additionally, we report the coverage rate, i.e., the fraction of \ac{ue} locations with at least
three valid \ac{toa} measurements for \ac{tdoa} localization. The \ac{mae} is computed only over
positions where the \ac{ls} solver converges to a finite estimate. To ensure statistical robustness, we report 95\% bootstrap \acp{ci} ($B = 1000$):
for \ac{marl}, the mean over the random seeds with a bootstrap \ac{ci} over the per-seed \ac{mae} values and
for the deterministic \ac{gdop} baselines, a bootstrap \ac{ci} over the \ac{ue} positions.

\section{Results and Discussion}
    \label{sec:results}
    \begin{table}[t]
\centering
\caption{Localization performance across five campus segments ($N=5$ BSs). MARL values are the mean over three seeds with a 95\,\% seed-bootstrap CI. GDOP values are deterministic with a 95\,\% UE-bootstrap CI ($B=1000$). Best MAE per segment in bold.}
\label{tab:results}
\setlength{\tabcolsep}{4pt}
\begin{tabular}{l cc cc cc}
\toprule
 & \multicolumn{2}{c}{GDOP Mean} & \multicolumn{2}{c}{GDOP Median} & \multicolumn{2}{c}{MARL} \\
\cmidrule(lr){2-3}\cmidrule(lr){4-5}\cmidrule(lr){6-7}
Segment & MAE & Cov. & MAE & Cov. & MAE & Cov. \\
 & [m] & [\%] & [m] & [\%] & [m] & [\%] \\
\midrule
S4 & 53.6$\,\pm\,$2.3 & 81 & 54.5$\,\pm\,$3.0 & 66 & \textbf{52.2}$\,\pm\,$1.8 & 79 \\
S6 & 88.0$\,\pm\,$3.2 & 89 & \textbf{67.7}$\,\pm\,$2.3 & 93 & 88.6$\,\pm\,$1.9 & 88 \\
S9 & 70.1$\,\pm\,$2.9 & 98 & 114.3$\,\pm\,$5.6 & 53 & \textbf{69.0}$\,\pm\,$1.9 & 85 \\
S11 & 82.1$\,\pm\,$2.5 & 91 & 77.4$\,\pm\,$2.0 & 90 & \textbf{70.8}$\,\pm\,$1.7 & 85 \\
S14 & \textbf{44.4}$\,\pm\,$1.9 & 99 & 45.6$\,\pm\,$2.1 & 99 & 46.4$\,\pm\,$1.3 & 96 \\
\midrule
Avg. & 67.6 & 92 & 71.9 & 80 & 65.4 & 87 \\
\bottomrule
\end{tabular}
\end{table}

We compare the proposed \ac{marl} approach with the \ac{gdop} baselines across five evaluation segments and
discuss the results and limitations.

\subsection{Results}
\begin{figure}[t]
    \centering
    \includegraphics[width=\columnwidth]{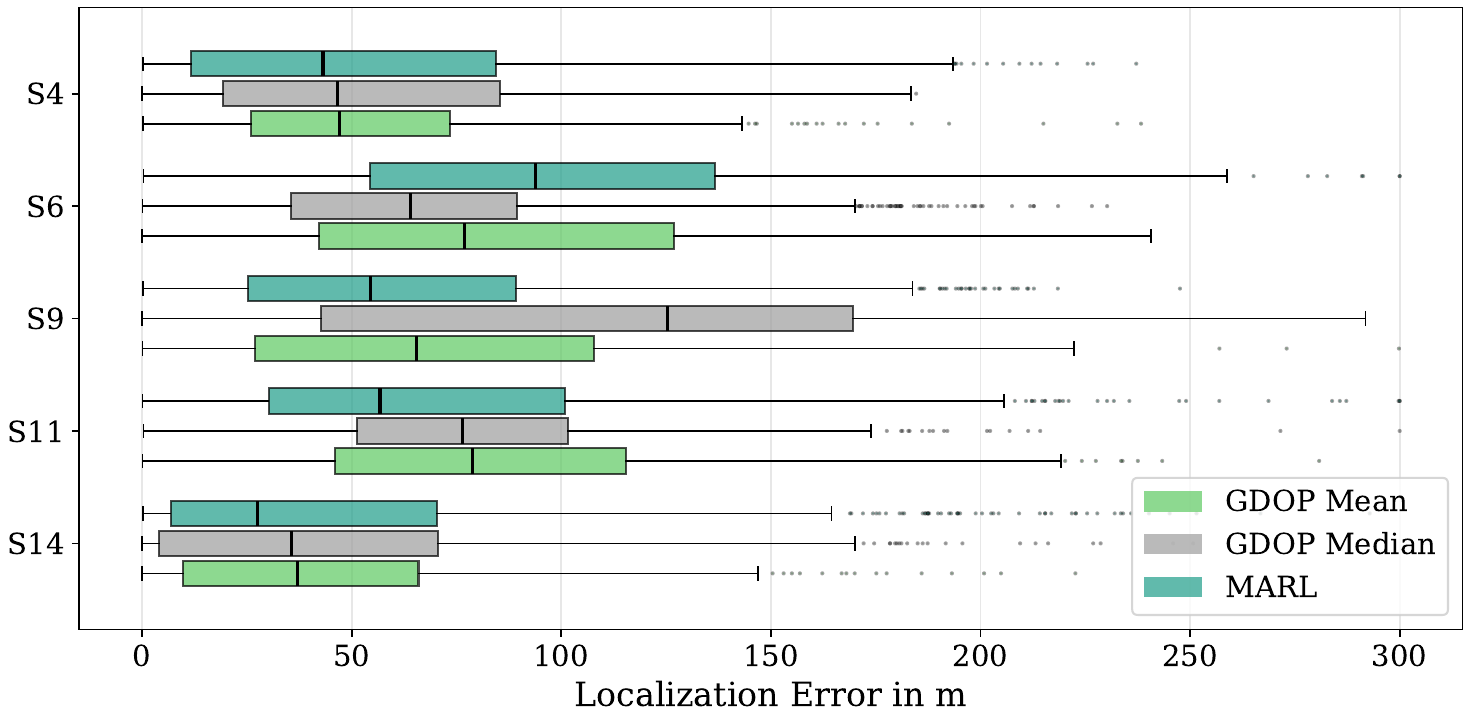}
    \caption{Distribution of localization error across all evaluation segments.}
    \label{fig:boxplots}
\end{figure}

Table~\ref{tab:results} summarizes the localization performance across all five evaluation segments,
none of which are used during training. On average, \ac{marl} achieves the lowest \ac{mae} of
\SI{65.4}{\meter}, compared with \SI{67.6}{\meter} for \ac{gdop} Mean and \SI{71.9}{\meter} for
\ac{gdop} Median, i.e.\ a reduction of about 3\,\% relative to the stronger (mean-optimized) and
about 9\,\% relative to the median-optimized geometric baseline.
\ac{marl} achieves the lowest \ac{mae} in three of the five segments (S4, S9, S11), but the
advantage is concentrated rather than uniform. It is by far largest in S11
(\SI{70.8}{\meter} vs.\ \SI{82.1}{\meter} for \ac{gdop} Mean, about 14\,\%), whereas the gains in
S4 and S9 are small (about 3\,\% and 2\,\% over \ac{gdop} Mean). In S6, \ac{gdop} Mean and \ac{marl} are
comparable around \SI{88}{\meter}, while \ac{gdop} Median is lowest at \SI{67.7}{\meter}, and in S14
\ac{gdop} Mean achieves the lowest \ac{mae} (\SI{44.4}{\meter} vs.\ \SI{46.4}{\meter} for \ac{marl}).
Regarding coverage, \ac{marl} localizes a smaller fraction of \acp{ue} than \ac{gdop} Mean on every
segment, and the gap is most notable in S9 (85\,\% vs.\ 98\,\%). Part of its \ac{mae} edge is therefore
computed over a more favorable subset of \acp{ue}, and the average comparison should be interpreted with this
coverage gap in mind.
Fig.~\ref{fig:boxplots} shows the per-segment error distributions and confirms the segment-dependent
pattern. \ac{marl} clearly reduces the median and spread in S11 and S9, is comparable in S4, and shows
a higher median in S6 and a heavier upper tail in S14.

\subsection{Discussion}
While \ac{marl} achieves the lowest average \ac{mae}, the advantage is concentrated in a single
segment rather than uniform across the scenario. Geometry-based \ac{gdop} treats all links between
\ac{bs} and \ac{ue} as equivalent and optimizes solely for geometric diversity. In contrast, the
\ac{marl} policy is trained on ray-tracing-derived \acp{cir} and can in principle learn to avoid \ac{bs}
positions that appear geometrically favorable but predominantly produce \ac{nlos}-biased \ac{toa} estimates.
The results, however, do not reveal a clear relationship between this potential benefit and segment
visibility. The largest improvement occurs in S11 (visibility 88.1\%), whereas the lowest-visibility
segments S4 and S9 (67.8\% and 74.3\%) yield only small gains, and in the most open segment S14
(93.3\%) \ac{gdop} Mean is slightly better. The benefit of propagation-aware placement is therefore
segment-specific and not explained by visibility alone. Attributing it to a single propagation
characteristic is not supported by the present evidence and is left for further investigation.

The absolute \ac{mae} values, on the order of \SIrange{45}{90}{\meter}, reflect the challenging scenario.
The limited number of $N=5$ \acp{bs} constrains geometric diversity, and \ac{nlos} links
introduce a systematic positive \ac{toa} bias that propagates into the \ac{tdoa} estimate.
Since no \ac{nlos} mitigation is applied, the focus is on relative improvement through
placement optimization rather than on minimizing absolute error.

\ac{marl} exhibits lower coverage than \ac{gdop} Mean on all five segments, most pronounced in S9
(85\% vs.\ 98\%). Because non-localizable \acp{ue} are excluded from the error computation, part
of the \ac{mae} difference may reflect this selection effect rather than placement quality alone.
The caveat is strongest precisely where \ac{marl}'s nominal \ac{mae} gain is small (S4 and S9), so it
tempers the average-\ac{mae} comparison. A coverage-matched evaluation would be needed to isolate the
placement contribution. The reduced coverage also suggests that the learned policy implicitly
prioritizes placements with more reliable propagation paths over maximum detectability when both
objectives are weighted equally.
Computationally, exhaustive search over all $\binom{M}{N}$ subsets is infeasible, which motivates both
the $\mathcal{O}(M)$ \ac{gdop} ranking and the learned policy, whose training is a one-time offline cost
amortized by a single forward pass per agent at inference.

Since all five evaluation segments are held out from training, the results indicate generalization to
unseen propagation conditions, though validation across more diverse environments is still needed.
Several limitations remain for future work, including the absence of estimator-level \ac{nlos}
mitigation and the restriction to a discrete candidate grid rather than continuous \ac{bs} positions.
\section{Conclusion}
    \label{sec:conclusion}
    In this paper, we proposed a \ac{marl} framework for optimizing \ac{bs} placement in \ac{tdoa} localization systems. 
By training on ray-tracing-generated \acp{cir} from a 3D model of a university campus, the approach 
implicitly learns to account for the effects of \acp{mpc} and \ac{nlos} conditions. 
Evaluation across five campus segments with varying propagation complexity shows
that the \ac{marl}-based optimization achieves localization accuracy comparable to conventional
\ac{gdop} baselines, lowering the average localization \ac{mae} by about 3\% relative to the
stronger geometric baseline, with a clear gain on one segment (S11, about 14\%) and comparable or
slightly higher error on the others. The advantage is thus segment-specific rather than uniform. On
some segments such as S14 the geometric baseline is slightly better, and \ac{marl} consistently
localizes fewer \acp{ue}, so the trade-off between accuracy and coverage warrants further investigation.
Future work will focus on validating the proposed approach with field measurements,
examining the impact of the number of \acp{bs} $N$ on placement quality,
and investigating generalization to environments with different propagation characteristics.

\section*{Acknowledgment}
This work was supported by the Bavarian Ministry of Economic Affairs, Regional Development and Energy (StMWi) within the project EMSIC (grant number: DIK0517/03). The work contributes to the research within the 6G-Valley innovation cluster. The authors acknowledge the use of AI-assisted tools for language refinement in the preparation of this manuscript. The authors alone are responsible for the content of this paper.

\bibliographystyle{IEEEtran}
\bibliography{references}

\end{document}